\documentclass{jpsj3}
\usepackage{txfonts}
\usepackage{braket}
\usepackage{color}
\usepackage{bm}
\usepackage{graphicx}
\usepackage{epstopdf}

\title{Emergent Anomalous Hall Effect in the Eu-Based Compound with a Diamond Network: The Centrosymmetric Cubic Antiferromagnet EuTi$_2$Al$_{20}$}

\author{Ryuji Higashinaka\thanks{E-mail address: higashin@tmu.ac.jp}$^{,\ 1}$, Kohsuke Sato$^{1}$, Ryosei Ideura$^{1}$, Masahiro Kawamata$^{1}$, and Tatsuma D. Matsuda$^{1}$}
\inst{$^{1}$Department of Physics, Tokyo Metropolitan University, Hachioji-shi, Tokyo 192-0397, Japan\\
}

\abst{
The centrosymmetric cubic compound EuTi$_2$Al$_{20}$, in which magnetic Eu ions form a diamond network, undergoes an antiferromagnetic transition at $T_{\rm N}=3.3$~K and exhibits metamagnetic transitions at $H_{\rm m1}=1.7$~T and $H_{\rm m2}=2.8$~T for $H\parallel[100]$ at 1.9~K. 
Between these fields, the magnetization shows a step-like behavior, defining an intermediate field-induced phase (Phase~II).
We investigated the electronic transport in Phase~II and found that both the resistivity and Hall resistivity are markedly enhanced, while remaining nearly field independent within the phase.
Phase~II appears for all field directions, although its transport response shows moderate directional dependence.
These features differ from the strongly orientation-selective behavior often observed in skyrmion-lattice phases of several $4f$-electron compounds, suggesting that Phase~II may host a field-induced spin texture with a topological character distinct from that of a conventional skyrmion lattice.
}
\date{\today}
\begin{document}
\maketitle
\section{Introduction}
Recently, the physical properties originating from the topological characteristics of electronic states in solids have emerged as a central focus in the study of strongly correlated electron systems.
These properties are characterized by quantized topological invariants such as the Chern number in the integer quantum Hall effect\cite{QuantumHall_82}, the $Z_2$ invariant in topological insulators\cite{TopoIns_05, TopoIns_07}, the winding number in topological superconductors\cite{TopoSC_00}, and the skyrmion number in magnetic skyrmions\cite{Skymion_16,Skymion_16-2}. 
Such topological quantities are robust against local perturbations including crystal defects and impurities.
Among these, systems exhibiting spin textures protected by topology have attracted particular interest as promising candidates for next-generation memory storage and information processing devices. 
A prototypical example is the magnetic skyrmion lattice (SkL), which possesses a vortex-like spin configuration\cite{Skyrmion_Review_21}.
In this system, the topologically protected spin vortex structure gives rise to an emergent magnetic field that acts on conduction electrons, inducing an additional Hall resistivity exclusively within the SkL phase\cite{Skyrmion_MnSi_09, THE_Gd2PdSi3_19}. 
This Hall resistivity cannot be explained by the conventional mechanisms such as the normal Hall effect, which is proportional to the external magnetic field ($H$), and/or the anomalous Hall effect, which scales with the bulk magnetization ($M$). 
Instead, it has been identified as the topological Hall effect.

The existence of a SkL was first experimentally confirmed in the B20-type compound MnSi, which possesses a chiral crystal structure lacking inversion symmetry\cite{Skyrmion_MnSi_09, Skyrmion_MnSi_09-2}. 
Initially, the formation of SkL state was considered to require Heisenberg spins with quenched orbital degrees of freedom, together with the Dzyaloshinskii-Moriya (DM) interaction, arising from the breaking of inversion symmetry, which stabilizes twisted magnetic structures.
However, beginning with the report of SkL formation in the chiral compound EuPtSi \cite{EuPtSi_18,EuPtSi_19}, where the Ruderman-Kittel-Kasuya-Yosida (RKKY) interaction constitutes the dominant exchange mechanism in a 4$f$-electron system, subsequent studies have also confirmed the presence of SkL formation and the topological Hall effect in centrosymmetric 4$f$-electron systems such as Gd$_2$PdSi$_3$ \cite{THE_Gd2PdSi3_19}, GdRu$_2X_2$ ($X$ = Si, Ge) \cite{GdRu2Si2_20, GdRu2Ge2_24}, as well as in the polar crystal EuNiGe$_3$ \cite{EuNiGe3_23, EuNiGe3_24}. 
These findings indicate that, in addition to the DM interaction, a variety of interactions—including, for example, magnetic frustration, RKKY interaction, and four-spin interactions—play a crucial role in the stabilization of SkL state.
The magnetic properties in these compounds arise from Eu$^{2+}$ and Gd$^{3+}$ ions, both of which exhibit Heisenberg spin behavior with no orbital degrees of freedom. 

In this study, we focus on EuTi$_2$Al$_{20}$, a centrosymmetric cubic antiferromagnet in which Eu$^{2+}$ ions carry the magnetic moment, as a new candidate material exhibiting SkL-like phase.
EuTi$_2$Al$_{20}$ belongs to the family of intermetallics with the general formula {\it Ln}{\it Tr}$_2X_{20}$ ({\it Ln}: rare-earth elements, {\it Tr}: transition metals, $X$: Zn, Al, Cd), which has attracted considerable interest in the field of strongly correlated electron systems due to its unique structural feature: the {\it Ln} sites are encapsulated within cages formed by sixteen $X$ atoms \cite{1-2-20_95, 1-2-20_98}. 
This cage structure provides a platform for investigating strongly correlated electronic behaviors of multipolar degrees of freedom associated with 4$f$ electrons.
Representative examples include the field-insensitive heavy fermion behavior observed in SmTi$_2$Al$_{20}$\cite{SmTi2Al20_11, SmTi2Al20_24} and the quadrupolar Kondo effect reported in Pr{\it Tr}$_2X_{20}$\cite{PrTr2Al20_11, PrTr2Zn20_11, PrNb_17}. 
Another important structural characteristic is that the {\it Ln} sites form a diamond network, which belongs to a non-symmorphic space group consisting of two face-centered cubic (fcc) sublattices displaced by (1/4, 1/4, 1/4), and possesses an inversion symmetry center located at the midpoint between adjacent sites of the sublattices.
It is known that strong magnetic frustration arises when the ratio of the nearest-neighbor interaction $J_1$ to the next-nearest-neighbor interaction $J_2$ satisfies $|J_2/J_1| \geq 1/8$\cite{Frustration_Diamond_07}. 
Previous studies have reported that EuTi$_2$Al$_{20}$ exhibits Curie-Weiss-like temperature ($T$) dependence of the magnetic susceptibility and undergoes an antiferromagnetic (AFM) transition at $T_{\rm N} = 3.23$ K\cite{EuTi2Al20_16}. 
The magnetization curve for magnetic field ($H$) applied along the [111] direction shows multi-step metamagnetic transitions at 1.5, 2.8, and 2.9 T, with a step-like behavior of magnetization observed in the intermediate field regions. 
As expected for Eu$^{2+}$ with $S = 7/2$ and $L = 0$, these multi-step metamagnetic transitions are observed at nearly the same field values irrespective of the field orientation, including for $H \parallel$ [100] and [110] directions. 
Although the effective magnetic moment ($\mu_{\rm eff}$) and the saturation magnetization ($M_{\rm sat}$) estimated from magnetization measurements are slightly smaller than the values expected for Eu$^{2+}$, the origin of this discrepancy remains unresolved.
Focusing on the the intermediate field region, we performed measurements of the electronic transport properties under magnetic fields applied in various directions to examine the possible realization of spin structures protected by topology.

\section{Experimental Details}
Single crystals of EuTi$_2$Al$_{20}$ were grown by the Al self-flux method at ambient pressure. 
Starting materials of 3N (99.9 \% pure) Eu grains, 4N Ti powders, and 4N Al grains were mixed in a molar ratio of 1:2:90, sealed in a quartz tube, and heated to 1100~$^\circ$C for 24~hr.
The mixture was rapidly cooled to 900~$^\circ$C and then cooled to 660~$^\circ$C at $-2~^\circ$C/hr. 
The excess Al was spun off in a centrifuge.
The single crystals are octahedron in shape bounded by \{111\} facets with a size of approximately 1.5 $\times$ 1.5 $\times$ 1.5 mm$^3$. 

Crystal structure refinement was performed at room temperature using a single-crystal X-ray diffractometer (XtaLAB mini, Rigaku) with graphite-monochromated Mo-K$\alpha$ radiation.
A selected small single crystal with dimensions of roughly 0.10 $\times$ 0.08 $\times$ 0.08 mm$^3$ was mounted on a glass fiber with epoxy. 
The structural parameters of EuTi$_2$Al$_{20}$ at room temperature refined using the SHELX-97 program\cite{SHELEX} are summarized in Table I. 
The reliability factors $R$ and $wR$ were smaller than the previous study ($wR$ = 10.8 \%\cite{EuTi2Al20_16}), which attest to the high quality of the crystal. 
Although possible Eu-site vacancies were discussed based on X-ray powder diffraction in Ref. [25], no significant site vacancies were detected in the present single-crystal refinement. 

The DC magnetic susceptibility $\chi$ and magnetization $M$ were measured in a Magnetic Property Measurement System 3 [MPMS3; Quantum Design (QD)] down to 1.9 K and up to 7 T. 
The actual magnitude of the applied magnetic field was estimated by taking into account the demagnetizing field arising from the sample geometry.
Specific heat ($C_p$) was measured by a quasi-adiabatic method with a Physical Property Measurement System (PPMS; QD) down to 2 K and up to 5 T. 
With a bar-shaped crystals with a size of $\sim$ 0.9 $\times$ 0.6 $\times$ 0.16 mm$^{3}$ with a (100) plane and a longest axis in the [011] direction, electrical resistivity ($\rho_{[011]}$) along [011] direction and Hall resistivity $\rho_{\rm H}$ were measured simultaneously using a Rotator and ACT option of the QD PPMS down to 1.9 K and up to 9 T.

\section{Results and Discussion}
%
We measured magnetization, resistivity, and specific heat using our single crystals and compared them with previous results \cite{EuTi2Al20_16}. 
Figure \ref{Tdep_HT}(a) shows $T$ dependence of $\chi$ and its inverse $\chi^{-1}$ along the $H \parallel$ [100] direction at $H$ = 0.1 T up to room temperature. 
The inset presents $T$ dependence of $\chi$ near the transition temperature under several fields applied along [100]. 
Below $T_{\rm N}$, $\chi$ decreases as expected for an AFM transition, and $T_{\rm N}$ shifts to lower temperature with increasing field. 
$\chi^{-1}$ displays a linear behavior over a wide temperature range from just above $T_{\rm N}$ to room temperature. 
A Curie-Weiss fit ($\chi^{-1} = \frac{T - \theta_{\rm W}}{C}$) between 50 and 300 K yields $C$ = 5.739(2) (emu$\cdot$K)/(mol$\cdot$Oe), corresponding to $\mu_{\rm eff} = 6.77\ \mu_{\rm B}$/Eu, and a Weiss temperature $\theta_{\rm W} = -1.26(5)$ K. 
The negative $\theta_{\rm W}$ indicates AFM correlations comparable in scale to $T_{\rm N}$.
The obtained $\mu_{\rm eff}$ is smaller than the theoretical value of 7.94$\mu_{\rm B}$/Eu for Eu$^{2+}$. 
A similar reduction in $\mu_{\rm eff}$ has also been reported in a previous study\cite{EuTi2Al20_16}. 
While that study attributed the reduced moment to the possible presence of Eu deficiency, our single-crystal structural analysis revealed no significant Eu-site vacancies.
Another possible explanation is the presence of mixed valence states of Eu. 
In materials, Eu can exist either as magnetic Eu$^{2+}$ with $S = 7/2$, or as nonmagnetic Eu$^{3+}$ with $J = 0$. 
Thus, if nonmagnetic Eu$^{3+}$ is partly present, the average valence exceeds 2+, resulting in a reduced effective magnetic moment. 
Indeed, according to $^{151}$Eu M\"ossbauer spectroscopy, the isomer shift of EuTi$_2$Al$_{20}$ at room temperature has been reported to be $-8.77(1)$~mm/s\,\cite{EuCr2Al20}. 
This value does not coincide with typical Eu$^{2+}$ ($\sim -10$ to $-12$~mm/s) or Eu$^{3+}$ ($\sim 0$~mm/s) values \cite{Eu8Ga16Ge30_24}, suggesting a possible valence deviation from the purely divalent state.
To clarify this possibility, experimental verification of the valence state of Eu is indispensable.

The resistivity $\rho_{[011]}$ measured along the $I\ \|$ [011] direction in Fig.~\ref{Tdep_HT}(b)  is metallic and exhibits a sharp drop below $T_{\rm N}$. 
As shown in the inset, $T_{\rm N}$ shifts to lower temperatures under $H\parallel[100]$, consistent with  $\chi(T)$. 
Furthermore, $\rho_{[011]}$ exhibits a minimum around 20~K and shows a weak upturn at lower temperatures. 
This behavior suggests possible contributions from either the magnetic scattering from a short-range ordering of Eu magnetic moments or the Kondo effect. 
A similar behavior related to the Kondo effect has been reported in Sm{\it Tr}$_2$Al$_{20}$ compounds \cite{SmTi2Al20_11, SmTa2Al20_13, SmTa2Al20_15}, where Sm shows an intermediate valence state. 
Although no magnetic susceptibility anomaly appears in the temperature range showing a $- \log T$ dependence of resistivity, dHvA measurements on SmTi$_2$Al$_{20}$ revealed an enhanced effective mass up to $26\,m_0$ \cite{SmTi2Al20_24}. 
These results suggest a possible unconventional Kondo effect arising from higher-order multipole moments and valence fluctuations, though definitive evidence is still lacking. 
Considering that Eu also has multiple 4$f$ electrons and possible intermediate valence, a similar unconventional Kondo effect cannot be excluded in the present compound.

Figure \ref{Tdep_HT}(c) presents the $T$ dependence of the magnetic specific heat divided by temperature, $C_{\rm mag}/T$, and the magnetic entropy, $S_{\rm mag}$. 
The non-magnetic contributions were subtracted using data from the non-magnetic analog LaTi$_2$Al$_{20}$. 
$C_{\rm mag}/T$ exhibits a sharp peak at $T_{\rm N}$, indicative of a first-order transition. 
Under $H\parallel[100]$, this sharp peak shifts to lower temperatures while maintaining its steepness.
The presence of latent heat near the transition temperature, as observed in the relaxation curves (data not shown), further supports the first-order nature of the transition. 
To estimate $S_{\rm mag}$ at the lowest measured temperature, a linear extrapolation of $C_{\rm mag}/T$ to zero at 0 K was assumed. 
Although $S_{\rm mag}$ does not reach the expected value of $R\ln 8$, which corresponds to the 4{\it f} electronic state with $S$ = 7/2, by 50 K, this discrepancy could be attributed either to an underestimation of $S_{\rm mag}$ due to the extrapolation and the presence of latent heat, or to possible valence fluctuations of the Eu ions.

Figure \ref{Tdep_LT} shows the $T$ dependence of $\chi$ at 0.1 T and $\rho_{[011]}$ and $C_p$ at 0 T in the vicinity of $T_{\rm N}$. 
The peak positions in the temperature derivatives of $\chi$ and $\rho_{[011]}$ coincide well with the $C_p$ peak, and we define $T_{\rm N}$ by these anomalies.  
No clear thermal hysteresis was detected in the measured quantities. 

Figure \ref{Hdep}(a) shows the $H$ dependence of the magnetization $M$ and its field derivative $dM/dH$ at 1.9 K for $H\parallel[100]$ up to 7 T, together with previously reported data for $H\ \|$ [111] up to 9 T\cite{EuTi2Al20_16}. 
Two metamagnetic transitions occur at $H_{\rm m1}$ = 1.7 T and $H_{\rm m2}$ = 2.8 T, between which $M(H)$ exhibits a step-like behavior.
An additional anomaly appears at $H_3\simeq 3.3$~T, suggesting the presence of another magnetic phase transition. 
These three magnetic transitions differ slightly from the values reported in the previous study \cite{EuTi2Al20_16}, which can be attributed to the difference in the direction of the applied magnetic field.
It has been confirmed that the transition fields reported in the previous study agree with those determined from the field dependence of the magnetoresistance (MR) measured under the same field orientation, as shown later. 
Moreover, $M$ at 7 T is approximately 5.5 $\mu_{\rm B}$/Eu, and the previously reported magnetization curve for $H \parallel$~[111] shown for comparison~\cite{EuTi2Al20_16} exhibits similarly good agreement with our results in the high-field region. 
Therefore, the magnetic anisotropy is considered to be negligible in high magnetic fields.
To precisely estimate the saturation magnetization, the magnetization curve in the PM phase at 4 K ($T > T_{\rm N}$) was self-consistently fitted using the Brillouin function that takes into account isotropic exchange interactions, as expressed below:
\begin{equation}
M = N g \mu_{\rm B} J B_J \left( \frac{g \mu_{\rm B} J (H + \lambda M)}{k_{\rm B} T} \right),
\end{equation}
\begin{equation}
B_J (x) = \frac{2J+1}{2J} \coth \left( \frac{2J+1}{2J} x \right) - \frac{1}{2J} \coth \left( \frac{x}{2J} \right), 
\end{equation}
where $N$, $g$, $k_{\rm B}$, $\lambda$, and $B_J(x)$ are number of spins, the g factor, the Boltzmann factor, the molecular field coefficient, and the Brillouin function, respectively. 
As a result, the saturation magnetization and molecular field coefficient were obtained as
$M_{\rm sat} = 5.59(9)\ \mu_{\rm B}$ and $\lambda = -0.36(6)$ T/$\mu_{\rm B}$, respectively.
This value of $M_{\rm sat}$ does not reach the theoretical value of $7~\mu_{\rm B}$ expected for Eu$^{2+}$. 
In the analysis of $\chi (T)$, several possible origins for the smaller magnetic moment than the theoretical value were proposed. 
In addition to those possibilities, it is also conceivable that the Eu magnetic moments are not yet fully saturated in this temperature and magnetic-field region, and that saturation may occur in a high-field phase above 10 T.  
To clarify this point, magnetization measurements in higher magnetic fields are indispensable.

Figure \ref{Hdep}(b, c) presents the $H$ dependence of $\rho_{[011]}$ and the Hall resistivity $\rho_{\rm H}$, simultaneously measured at 1.9 K for $H\parallel[100]$ and $I\parallel[011]$. 
In Phase I ($H < H_{\rm m1}$), both quantities increase and exhibit pronounced anomalies at  $H_{\rm m1}$. 
In the intermediate region $H_{\rm m1} < H < H_{\rm m2}$ (Phase II), $\rho_{[011]}$ is nearly field independent, while $\rho_{\rm H}$ increases linearly with $H$. 
As the field approaches $H_{\rm m2}$, both quantities undergo a steep decrease.
The $H$-independent behavior of $\rho_{[011]}$ in Phase II, in conjunction with the presence of a step-like behavior of magnetization, suggests the absence of magnetic structure changes within this phase. 
In the high-field region $H_{\rm m2} < H < H_3$ (Phase III), $\rho_{[011]}$ increases while $\rho_{\rm H}$ decreases. 
At $H = H_3$, $\rho_{[011]}$ shows a peak and subsequently decreases, whereas $\rho_{\rm H}$ exhibits an anomaly resembling a shoulder-like feature.
No hysteresis was observed with respect to the magnetic field for any of these features. 
The prominent enhancements in MR and Hall effect observed in the intermediate magnetic phase (Phase II) are reminiscent of those reported in various SkL states, which are characterized by spin textures with discrete topological numbers.\cite{Skyrmion_MnSi_09,THE_Gd2PdSi3_19,EuPtSi_18,GdRu2Si2_20,GdRu2Ge2_24,EuNiGe3_20}

In materials exhibiting the SkL phase, the Hall resistivity can be expressed as the sum of three distinct contributions: the normal Hall term $R_0 H$, the anomalous Hall term $R_s M$ proportional to the magnetization, and the topological Hall term $\rho^{\rm T}_{\rm H}$ arising from the emergent magnetic field induced by the spin structure protected by topology.
Regarding the anomalous Hall contribution, several mechanisms have been proposed, including scattering mechanisms such as skew scattering\cite{Smit_1955} and side-jump scattering \cite{Berger_1964} and intrinsic mechanisms derived from the band structure and Berry curvature\cite{Karplus_Luttinger_1954}. 
However, regardless of the model employed, it was not possible to satisfactorily fit the entire field-dependent Hall resistivity using only the normal and anomalous Hall terms.
These findings suggest that, in phases II and III, additional topological terms beyond the conventional and anomalous Hall contributions may be present. 
Within these phases, there is a possibility of emergent ordered states characterized by specific topological numbers, including those associated with SkL, wherein emergent magnetic fields may arise.

Figure \ref{100} presents the $H$ dependence of $\rho_{[011]}$ and $\rho_{\rm H}$ at 1.9 to 5 K for $H \parallel$ [100], along with the $H$-$T$ phase diagram determined in the present study, shown with a color plot of $\rho_{\rm H}$. 
With increasing temperature, the MR retains its weak $H$ dependence in Phase II, while the three phase transitions shift toward lower magnetic fields and eventually merge. 
All associated anomalies disappear around 3.5 K. 
The enhanced-$\rho_{\rm H}$ region associated with Phase~II shrinks with increasing temperature but remains open down to 1.9~K, so measurements at lower temperatures are needed to complete the phase diagram.
A notable feature is that $\rho_{\rm H}$ within Phase~II is nearly temperature independent, in contrast to many SkL systems where the topological Hall contribution decreases with temperature \cite{Skyrmion_MnSi_09,THE_Gd2PdSi3_19,EuPtSi_18,GdRu2Si2_20,GdRu2Ge2_24,EuNiGe3_20}.

Figure~\ref{110_111} compares MR at 1.9 K for $H\parallel[100]$, $[0\bar{1}1]$, and $[1\bar{1}1]$ and shows the corresponding phase diagrams.
These measurements were performed on the same sample by rotating it to change the field orientation, as illustrated in the figure. 
No pronounced anisotropy is observed within Phase I, whereas in Phase II, an anisotropic behavior appears, exhibiting a weak field dependence but different resistivity values along distinct directions.
Upon further application of the magnetic field, for $H \parallel$ [0$\bar{1}$1] and [1$\bar{1}$1], the MR increases, exhibiting an anomaly indicative of the narrow region of  phase III. 
Subsequently, as the system enters PM phase, the MR decreases, reaching a minimum around 6 T, followed by an upward trend.
In contrast, for $H \parallel$ [100], the MR exhibits a sharp decrease upon entering phase III, followed by a peak at the phase boundary with PM phase, and then shows behavior within PM phase similar to that observed for other magnetic field directions.
A comparison of the $H$-$T$ phase diagrams for the three field orientations reveals that the phase II region is widest for $H \parallel$ [1${\bar{1}}$1], indicating that phase II is most stabilized in this direction. 
On the other hand, phase III extends to higher magnetic fields and occupies the largest region when $H \parallel$ [100].
These results indicate that in the present compound, the anisotropy of the $H$-$T$ phase diagram is not significant, and that Phase II, in which an additional Hall effect was observed, exists irrespective of the magnetic field orientation.

Figure~\ref{Angledep} shows the angular dependence of the normalized $\rho_{\rm H}$ measured while rotating $H$ from [100] to [0$\bar{1}$1] for $I\parallel[011]$ in phase~II ($T = 1.9$~K, $H = 2.2$~T) and PM phase ($T = 5$~K, $H = 9$~T).
In PM phase, $\rho_{\rm H}$ follows a cosine-like angular dependence, consistent with the normal Hall response proportional to the [100] component of the magnetic field. 
In Phase~II, the angular dependence deviates from a pure cosine curve and shows an anomalous sign reversal when the field approaches the voltage-measurement direction ($H\parallel[0\bar{1}1]$).
The origin of this sign reversal is presently unclear, and further work is needed to determine whether it is intrinsic or caused by misalignment or other extrinsic effects.
No hysteresis was observed with respect to the direction of magnetic field rotation in either case. 
For comparison, centrosymmetric hexagonal Gd$_2$PdSi$_3$ exhibits a rectangular-like angular dependence with strong anisotropy and hysteresis due to crystal anisotropy \cite{THE_Gd2PdSi3_19}, which is qualitatively different from the present behavior. 

To understand the behavior of the additional Hall effect observed in phase II of this material, we compare it with other systems that exhibit an enhancement of the Hall effect associated with the formation of a SkL phase.  
Among cubic compounds exhibiting an enhanced Hall effect in the intermediate field region, chiral crystals such as MnSi \cite{Skyrmion_MnSi_09} and EuPtSi \cite{EuPtSi_18} are well known.  
In these systems, the formation of the SkL state induces a topological Hall contribution.
In MnSi, the magnetic moment is carried by the {\it d} electrons, and the skyrmion size is as large as 180~\AA, corresponding to about forty times the lattice constant.  
The field-induced intermediate phase (A phase) exists for all directions of the applied magnetic field \cite{Skyrmion_MnSi_09-2}.  
Such a long-period magnetic structure is considered to reflect the ratio between the dominant exchange interaction and the DM interaction among the {\it d} electrons, namely $J_c / J_{\rm DM}$.
In contrast, in EuPtSi, where the magnetic moment arises from the {\it f} electrons, the skyrmion size is much smaller—about 18~\AA, corresponding to four times the lattice constant—and the SkL phase has been observed only for $H \parallel [111]$ \cite{EuPtSi_19, EuPtSi_24}.  
For $H \parallel [100]$, the SkL phase has not been clearly identified, and for $H \parallel [110]$ it is reported to be absent \cite{EuPtSi_19-2}, indicating strong anisotropic behavior.  
This difference in skyrmion size likely reflects the nature of the dominant magnetic interactions: the exchange interaction in the {\it d}-electron systems, and the conduction-electron-mediated RKKY interaction in the {\it f}-electron systems.

Furthermore, when the magnetic modulation period is long, the magnetic structure can be treated within the continuum approximation with respect to the underlying lattice \cite{Skyrmion_MnSi_09-2}.  
In contrast, for shorter modulation periods, stronger coupling between the magnetic structure and the lattice is expected.  
As a result, the spin texture becomes more unstable against the direction of the applied magnetic field, which can lead to a highly anisotropic $H$-$T$ phase diagram \cite{EuPtSi_19-2}.
Indeed, in several Gd-based compounds that exhibit localized spin magnetism similar to that of Eu$^{2+}$, SkL phases with comparable short modulation periods have been reported \cite{THE_Gd2PdSi3_19}, and these phases also show a strong dependence on the applied magnetic field direction.

SkL phases in $4f$-electron systems, such as EuPtSi, generally exhibit strong anisotropy with respect to the magnetic-field direction.
Accordingly, the intermediate phase (Phase~II) of EuTi$_2$Al$_{20}$, in which magnetism is also governed by $4f$ electrons, is expected to show an anisotropic $H$--$T$ phase diagram.
Contrary to this expectation, Phase~II is stably realized for all field directions and occupies a markedly wider region than the SkL phases reported in other $4f$-electron compounds.
In addition, the additional anomalous Hall term observed in the intermediate-field phase exhibits almost no temperature dependence, which is distinct from the behavior of SkL phases in previously studied $4f$-electron systems.
These differences suggest that the magnetic structure realized in Phase~II of EuTi$_2$Al$_{20}$ is likely an ordered state characterized by a topological spin number different from that of conventional SkL states in other $4f$-electron compounds.
To clarify this possibility, it is essential to elucidate the magnetic structures within the magnetically ordered phases of this material.

Next, we consider a comparison with MnSc$_2$S$_4$, an $A$-site spinel compound in which Mn$^{2+}$ magnetic ions—possessing Heisenberg spin characteristics similar to those in the present material—form a diamond network, although it is an electrical insulator. 
This compound is known to exhibit strong magnetic frustration, characterized by a ratio of $|J_2/J_1| = 0.85$. 
Reflecting this frustration, neutron scattering experiments have revealed a variety of magnetic ordering phases. 
Under zero magnetic field, MnSc$_2$S$_4$ exhibits successive transitions from a commensurate spin-density wave (SDW) to an incommensurate SDW phase, eventually developing a helical magnetic order characterized by a single propagation vector $\bm{q} = (0.75, 0.75, 0)$ at the lowest temperatures. 
Upon applying a magnetic field along the [001] direction, an intermediate magnetic phase characterized by a triple-$\bm{q}$ structure emerges\cite{MnSc2S4_17}. 
Within this phase, the magnon thermal Hall effect has been observed, and the realization of a vortex-like state protected by topology—analogous to a SkL phase—has been proposed \cite{MnSc2S4_24}. 
For EuTi$_2$Al$_{20}$, single-crystal neutron diffraction studies under zero field reported magnetic reflections at (110), consistent with a collinear order with propagation along (001) \cite{EuTi2Al20_Neu_21}, and our recent powder neutron diffraction and resonant X-ray scattering results are consistent with this picture \cite{Kawamata_25}.
In contrast to the relatively simple (001)-type collinear magnetic structure proposed for the present material, materials such as MnSc$_2$S$_4$ that host SkL-like phases typically exhibit complex magnetic structures—such as helical or incommensurate orders—adjacent to SkL phases. 
As of now, the magnetic structures of Phases II and III under applied magnetic fields in EuTi$_2$Al$_{20}$ remain experimentally uncharacterized. 
Further investigations using neutron and resonant X-ray scattering techniques in these field-induced phases are therefore necessary.
\section{Summary}
We investigated magnetotransport in EuTi$_2$Al$_{20}$, a centrosymmetric cubic antiferromagnet with a diamond network of Eu moments.
For $H\parallel[100]$ at 1.9~K, metamagnetic transitions occur at $H_{\rm m1}$ and $H_{\rm m2}$, and an intermediate phase (Phase~II) is identified between them.
Within Phase~II, both the magnetoresistance and Hall resistivity are strongly enhanced and exhibit only weak field dependence; moreover, the Hall resistivity in Phase~II is nearly temperature independent.
Phase~II exists for all field directions, indicating robust stabilization against field rotation.
These behaviors differ from those typically observed in SkL phases of $4f$-electron compounds, suggesting a distinct topological spin texture in EuTi$_2$Al$_{20}$.
Direct determination of the magnetic structures in the field-induced phases is crucial for clarifying the origin of the additional Hall response.
\begin{acknowledgments}
We thank Yoshikazu Mizuguchi and Aichi Yamashita for the magnetization measurements, Yoshichika \={O}nuki, Kazumasa Hattori, and Satoshi Tsutsui for fruitful discussions, and Yuji Aoki for discussions and maintenance of the equipments. 
This work was supported by MEXT/JSPS KAKENHI Grants Number JP25K07228, JP23H04866, JP23H04870, JP22K03517, and JP22K03522. 
\end{acknowledgments}

\clearpage
\begin{figure}
\caption{(Color online) 
$T$ dependence of (a) magnetic susceptibility $\chi$ at 0.1 T and its inverse, (b) resistivity $\rho_{[011]}$, and (c) magnetic specific heat divided by temperature $C_{\rm mag}/T$ and magnetic entropy $S_{\rm mag}$ at 0 T. 
Insets show the data near $T_{\rm N}$ under fields along [100].  
In (c), the phonon contribution was subtracted using LaTi$_2$Al$_{20}$ as a nonmagnetic reference; $S_{\rm mag}$ was obtained by integrating $C_{\rm mag}/T$ with a linear extrapolation of $C_{\rm mag}/T\to 0$ at $T\to 0$. 
}
\label{Tdep_HT}
\end{figure}
\begin{figure}
\caption{(Color online) 
$T$ dependence near $T_{\rm N}$ of (a) $\chi$ at 0.1 T and (b) $\rho_{[011]}$, and (c) $C_p$ at 0 T. 
Peaks in $d\chi/dT$ and $d\rho_{[011]}/dT$ coincide with the peak in $C_p$, which is used to define $T_{\rm N}$. 
}
\label{Tdep_LT}
\end{figure}
\begin{figure}
\caption{(Color online) 
$H$ dependence at $T$ = 1.9 K for $H\ \|$ [100] of (a) magnetization $M$ and its field derivative, (b) $\rho_{[011]}$, and (c) Hall resistivity $\rho_{\rm H}$.
In (a), the magnetization curve for $H \parallel$ [111] up to 9 T, as reported in a previous study\cite{EuTi2Al20_16}, is additionally shown as a dotted line.
The characteristic fields $H_{\rm m1}$, $H_{\rm m2}$ and $H_3$ are defined as the peaks in $dM/dH$.
Further details are provided in the main text.
}
\label{Hdep}
\end{figure}
\begin{figure}
\caption{(Color online) 
$H$ dependence of (a) $\rho_{[011]}$ and (b) $\rho_{\rm H}$ for $H\ \|$ [100] at various temperatures, and (c) the resulting $H$-$T$ phase diagram along with a color map of $\rho_{\rm H}$. 
Phase II shifts to lower fields and its area diminishes as the temperature increases, eventually disappearing at the boundary with PM phase.
Within Phase II, the values of $\rho_{[011]}$ and Hall resistivity divided by the magnetic field $\rho_{\rm H}/H$ (i.e., the Hall coefficient) remain almost constant.
}
\label{100}
\end{figure}
\begin{figure}
\caption{(Color online) 
(a) $H$ dependence of $\rho_{[011]}$ at $T$ = 1.9 K for $H\ \|$ [100], [0$\bar{1}$1], and [1$\bar{1}$1],  and (b, c) corresponding $H$-$T$ phase diagrams for $H\ \|$ [0$\bar{1}$1] and [1$\bar{1}$1].
In phase I, magnetic anisotropy is negligible; however, clear anisotropic behavior is observed in phases II and III. 
Phase~II appears for all directions and is widest for $H\parallel[1\bar{1}1]$. 
Phase~III extends to higher fields for $H\parallel[100]$. 
}
\label{110_111}
\end{figure}

\begin{figure}[thb]
\caption{(Color online) 
Angular dependence of the normalized $\rho_{\rm H}$ measured while rotating $H$ from [100] to $[0\bar{1}1]$ for $I\parallel[011]$, in Phase~II (1.9~K, 2.2~T) and in PM phase (5~K, 9~T).  
The relationship between the directions of current, voltage measurements, and the applied magnetic field is illustrated in the figure. 
In PM phase, $\rho_{\rm H}$ follows a cosine-like dependence (solid curve), consistent with the normal Hall effect.
In Phase~II, the angular profile deviates from a pure cosine and shows a sign reversal near $H\parallel[0\bar{1}1]$.
No hysteresis with respect to the rotation direction was observed in either set of data.  
}
\label{Angledep}
\end{figure}
\clearpage

\begin{table*}[tb]
\caption{Atomic coordinates and thermal parameters of EuTi$_2$Al$_{20}$ at room temperature determined by single-crystal X-ray measurements. 
$Z$ is the number of formula units in a crystallographic unit cell. 
$R$ and $wR$ are the reliability factors, $B_{\rm eq}$ is the equivalent isotropic atomic displacement parameter, and Occ. is the occupancy of each site. 
Standard deviations in the positions of the least significant digits are given in parentheses.}
\label{t1}
\begin{center}
\begin{tabular}{lrlccccl}
\hline
\multicolumn{3}{c}{$Fd{\bar 3}m$, $O_{h}^{6}$ ($\sharp$227)} & \multicolumn{4}{c}{$a$ $=$ 14.7257(13) $\AA$, $V$ $= $ 3193.2(5) $\AA^3$}\\
\multicolumn{3}{c}{$Z$ = 8} & \multicolumn{3}{c}{Position}\\
 \cline{4-6}
Atom & Site & & $x$ & $y$ & $z$ & $B_{\rm eq}\ (\AA^2)$ & Occ. \\
\hline
Al(1) & $96g$ & ($..m$) & 0.05949(6) & 0.05949(6) & 0.32502(8) & 0.850(12) & 0.990(26) \\
Al(2) & $48f$ & ($2.mm$) & 0.48663(11) & 1/8 & 1/8 & 0.623(14) & 0.988(24) \\
Ti & $16d$ & ($.{\bar 3}m$) & 1/2 & 1/2 & 1/2 & 0.396(13) & 0.996(24) \\
Al(3) & $16c$ & ($.{\bar 3}m$) & 0 & 0 & 0 & 1.56(2) & 1.008(24) \\
Eu & $8a$ & (${\bar 4}3m$) & 1/8 & 1/8 & 1/8 & 0.636(10) & 0.998(26)\\
\hline
\multicolumn{3}{c}{$R$ $=$ 2.14$\%$, $wR$ $=$ 5.73$\%$}
\end{tabular}
\end{center}
\label{t1}
\end{table*}

\clearpage

\setcounter{figure}{0}
\begin{figure}[thb]
\begin{center}
\includegraphics[width=0.8\linewidth]{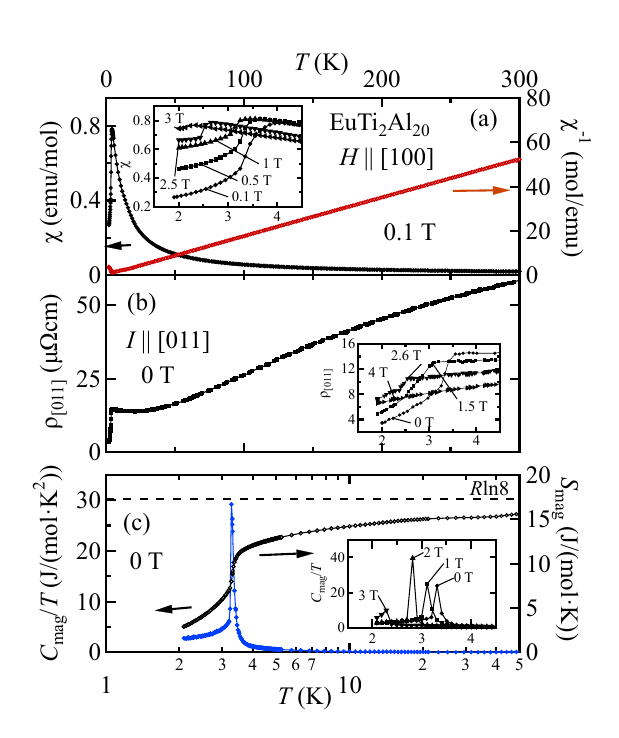}
\end{center}
\caption{}
\label{Tdep_HT}
\end{figure}
\begin{figure}
\begin{center}
\includegraphics[width=0.7\linewidth]{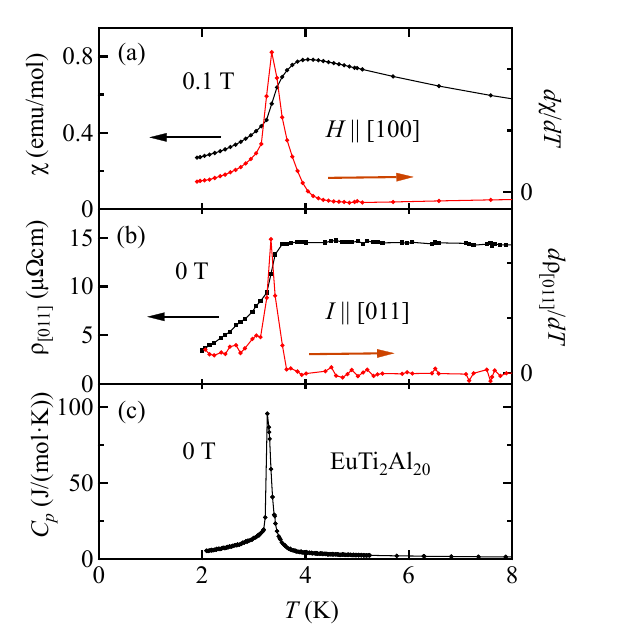}
\end{center}
\caption{}
\label{Tdep_LT}
\end{figure}
\begin{figure}
\begin{center}
\includegraphics[width=0.7\linewidth]{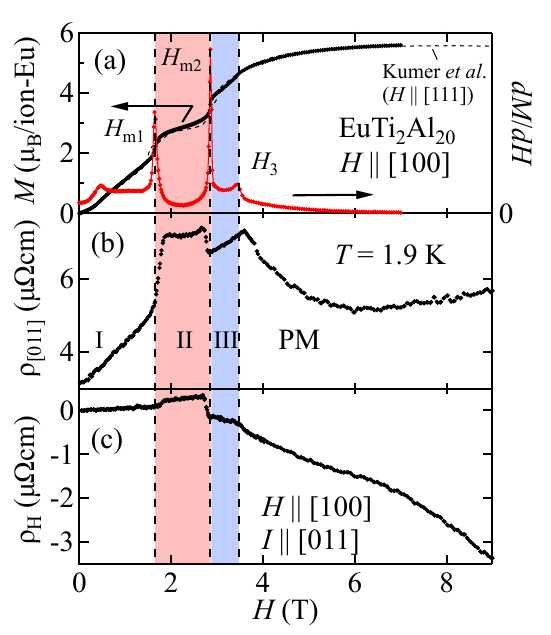}
\end{center}
\caption{}
\label{Hdep}
\end{figure}
\begin{figure}
\begin{center}
\includegraphics[width=\linewidth]{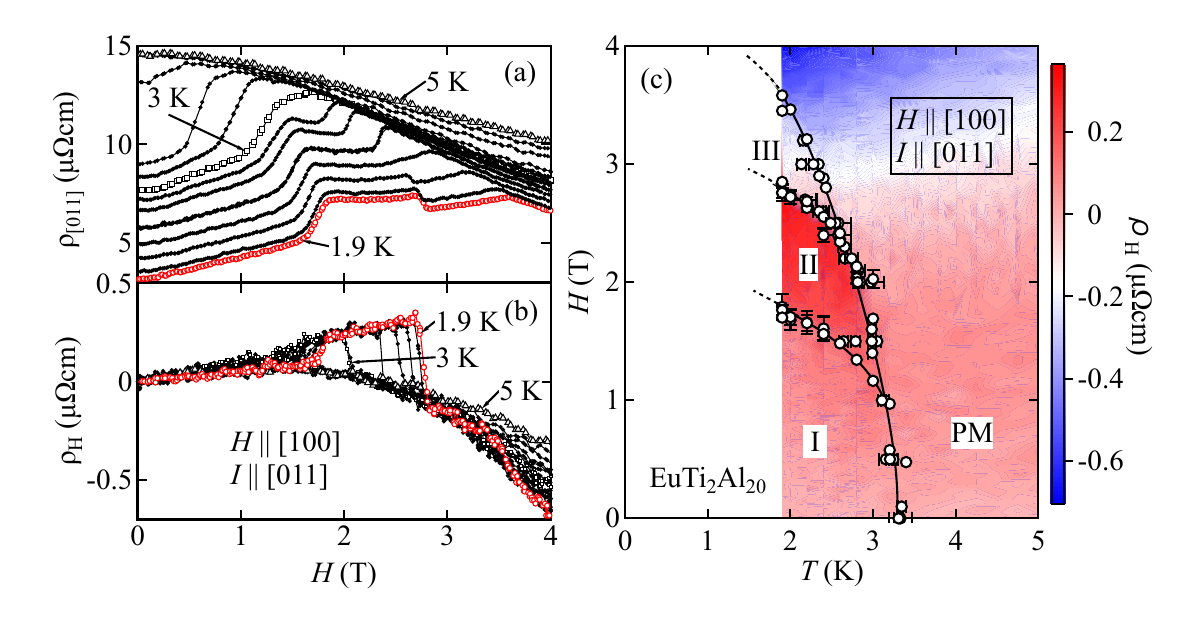}
\end{center}
\caption{}
\label{100}
\end{figure}
\begin{figure}
\begin{center}
\includegraphics[width=\linewidth]{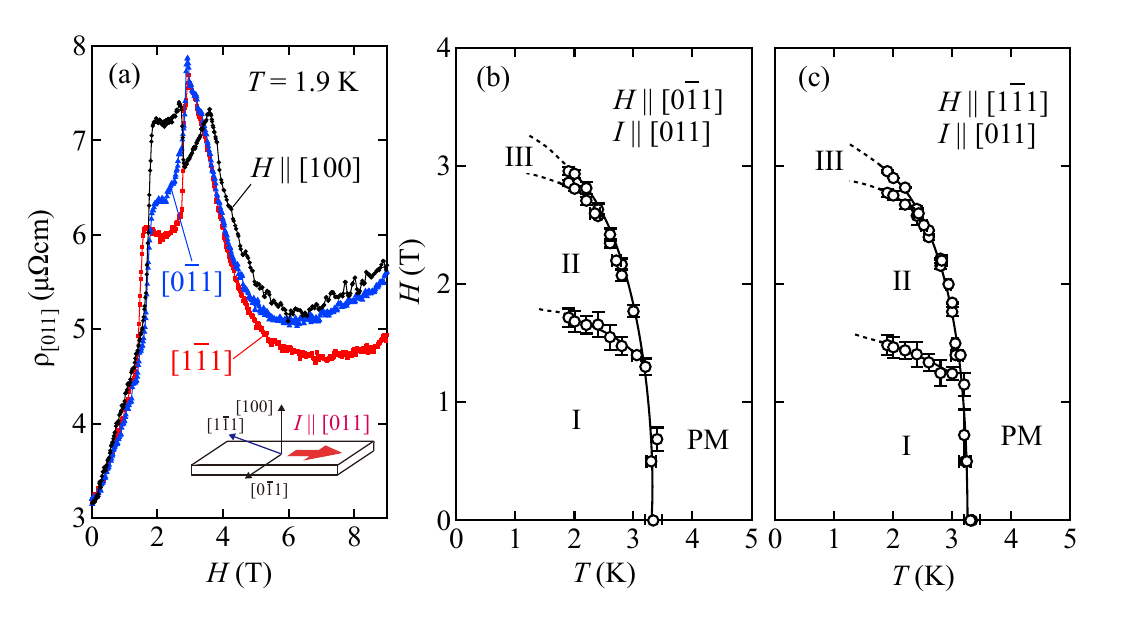}
\end{center}
\caption{}
\label{110_111}
\end{figure}
\begin{figure}
\begin{center}
\includegraphics[width=0.7\linewidth]{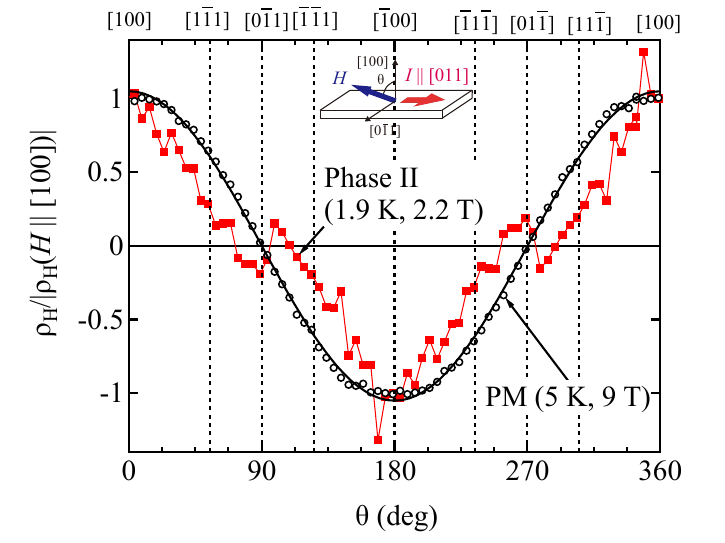}
\end{center}
\caption{}
\label{Angledep}
\end{figure}

\clearpage
\begin{thebibliography}{99}
\bibitem{QuantumHall_82} D. J. Thouless, M. Kohmoto, M. P. Nightingale, and M. den Nijs, Phys. Rev. Lett. {\bf 49}, 405 (1982).
\bibitem{TopoIns_05} C. L. Kane and E. J. Mele, Phys. Rev. Lett. {\bf 95}, 146802 (2005).
\bibitem{TopoIns_07} L. Fu and C. L. Kane, Phys. Rev. B {\bf 76}, 045302 (2007).
\bibitem{TopoSC_00} N. Read and D. Green, Phys. Rev. B {\bf 61}, 10267 (2000).
\bibitem{Skymion_16} M. B. A. Jalil, S. G. Tan, Z. B. Siu, W. Gan, I. Purnama, and W. S. Lew, J. Magn. Mag. Mater. {\bf 399}, 155 (2016).
\bibitem{Skymion_16-2} X. Zhang, Y. Zhou, and M. Ezawa, Phys. Rev. B {\bf 93}, 024415 (2016).
\bibitem{Skyrmion_Review_21} Y. Tokura and N. Kanazawa, Chem. Rev. {\bf 121}, 2857 (2021).
\bibitem{Skyrmion_MnSi_09} A. Neubauer, C. Pfleiderer, B. Binz, A. Rosch, R. Ritz, P. G. Niklowitz, and P. B\"{o}ni, Phys. Rev. Lett. {\bf 102}, 186602 (2009).
\bibitem{THE_Gd2PdSi3_19} T. Kurumaji, T. Nakajima, M. Hirschberger, A. Kikkawa, Y. Yamasaki, H. Sagayama, H. Nakao, Y. Taguchi, T. Arima, and Y. Tokura, Science {\bf 365}, 914 (2019).
\bibitem{Skyrmion_MnSi_09-2} S. M\"{u}hlbauer, B. Binz, F. Jonietz, C. Pfleiderer, A. Rosch, A. Neubauer, R. Georgii, and P. B\"{o}ni, Science {\bf 323} 915 (2009).

\bibitem{EuPtSi_18} M. Kakihana, D. Aoki, A. Nakamura, F. Honda, M. Nakashima, Y. Amako, S. Nakamura, T. Sakakibara, M. Hedo, T. Nakama, and Y. \={O}nuki, J. Phys. Soc. Jpn. {\bf 87}, 023701 (2018).
\bibitem{EuPtSi_19} K. Kaneko, M. D. Frontzek, M. Matsuda, A. Nakao, K. Munakata, T. Ohhara, M. Kakihana, Y. Haga, M. Hedo, T. Nakama, and Y. \={O}nuki, J. Phys. Soc. Jpn. {\bf 88}, 013702 (2019).
\bibitem{GdRu2Si2_20} N. D. Khanh, T. Nakajima, X. Yu, S. Gao, K. Shibata, M. Hirschberger, Y. Yamasaki, H. Sagayama, H. Nakao, L. Peng, K. Nakajima, R. Takagi, T. Arima, Y. Tokura and S. Seki, Nat. Nanotechnol. {\bf 15} 444 (2020).
\bibitem{GdRu2Ge2_24} H. Yoshimochi, R. Takagi, J. Ju, N. D. Khanh, H. Saito, H. Sagayama, H. Nakao, S. Itoh, Y. Tokura, T. Arima, S. Hayami, T. Nakajima and S. Seki, Nat. Phys. {\bf 20} 1001 (2024).
\bibitem{EuNiGe3_23} D. Singh, Y. Fujishiro, S. Hayami, S. H. Moody, T. Nomoto, P. R. Baral, V. Ukleev, R. Cubitt, N-J. Steinke, D. J. Gawryluk, E. Pomjakushina, Y. \={O}nuki, R. Arita, Y. Tokura, N. Kanazawa, and J. S. White, Nat. Comm. {\bf 14}, 8050 (2023).
\bibitem{EuNiGe3_24} T. Matsumura, K. Kurauchi, M. Tsukagoshi, N. Higa, H. Nakao, M. Kakihana, M. Hedo, T. Nakama, and Y. \={O}nuki, J. Phys. Soc. Jpn. {\bf 93}, 074705 (2024).
\bibitem{1-2-20_95} S. Niemann and W. Jeitschko, J. Solid State Chem. {\bf 114}, 337 (1995).
\bibitem{1-2-20_98} V. M. T. Thiede, W. Jeitschko, S. Niemann, and T. Ebel, J. Alloys Comp. {\bf 67}, 23 (1998). 
\bibitem{SmTi2Al20_11} R. Higashinaka, T. Maruyama, A. Nakama, R. Miyazaki, Y. Aoki, and H. Sato, J. Phys. Soc. Jpn. {\bf 80}, 093703 (2011).
\bibitem{SmTi2Al20_24} Md A. Afzal, Y. \_{O}nuki, D. Aoki, H. Harima, R. Higashinaka, Y. Aoki, and T. D. Matsuda, J. Phys. Soc. Jpn. {\bf 93}, 054710 (2024).
\bibitem{PrTr2Al20_11} A. Sakai and S. Nakatsuji, J. Phys. Soc. Jpn. {\bf 80}, 063701 (2011).
\bibitem{PrTr2Zn20_11} T. Onimaru, K. T. Matsumoto, Y. F. Inoue, K. Umeo, T. Sakakibara, Y. Karaki, M. Kubota, and T. Takabatake, Phys. Rev. Lett. {\bf 106}, 77001 (2011).
\bibitem{PrNb_17} R. Higashinaka, A. Nakama, R. Miyazaki, J. Yamaura, H. Sato, and Y. Aoki, J. Phys. Soc. Jpn. {\bf 86}, 103703 (2017).
\bibitem{Frustration_Diamond_07} D. Bergman, J. Alicea, E. Gull, S. Trebst, and L. Balents, Nat. Phys. {\bf 3}, 487 (2007).
\bibitem{EuTi2Al20_16} R. Kumar, H. S. Nair, R. Christian, A. Thamizhavel, and A. M. Strydom, J. Phys. Mater. {\bf 28} 436002  (2016).
\bibitem{SHELEX} G. M. Sheldrick, SHELX-97: Program for the Solution for Crystal Structures. University of G\"ottingen, Germany, 1997.
\bibitem{EuCr2Al20} A. Koldemir, S. Klenner and R. P\"ottgen, Z. Kristallogr. - N. Cryst. Struct. {\bf 268(3)}, 507 (2023).
\bibitem{Eu8Ga16Ge30_24} S. Tsutsui, Y. Kobayashi, M. Mizumaki, N. Kawamura, M. K. Kubo, S. Ikeda, H. Kobayashi, Y. Yoda, T. Onimaru, M. A. Avila, and T. Takabatake, J. Phys. Soc. Jpn. {\bf 93}, 084702 (2024).
\bibitem{SmTa2Al20_13} A. Yamada, R. Higashinaka, R. Miyazaki, K. Fushiya, T. D. Matsuda, Y. Aoki, W. Fujita, H. Harima, and H. Sato, J. Phys. Soc. Jpn. {\bf 82}, 123710 (2013).
\bibitem{SmTa2Al20_15} A. Yamada, R. Higashinaka, T. D. Matsuda, Y. Aoki, and H. Sato, J. Phys. Soc. Jpn. {\bf 84}, 103701 (2015).
\bibitem{EuNiGe3_20} W. Iha, S. Matsuda, M. Kakihana, D. Aoki, A. Nakamura, M. Nakashima, Y. Amako, T. Takeuchi, M. Kimata, Y. Otani, M. Hedo, T. Nakama, and Y. \={O}nuki, JPS Conf. Proc. {\bf 30}, 011092 (2020).
\bibitem{Smit_1955} J. Smit, Physica (Amsterdam) {\bf 21}, 877 (1954).
\bibitem{Berger_1964} L. Berger, Physica (Amsterdam) {\bf 30}, 1141 (1964).
\bibitem{Karplus_Luttinger_1954} R. Karplus and J. M. Luttinger, Phys. Rev. {\bf 95}, 1154 (1954).
\bibitem{EuPtSi_24}
T. Matsumura, C. Tabata, K. Kaneko, H. Nakao, M. Kakihana, M. Hedo, T. Nakama, and Y. \={O}nuki, Phys. Rev. B {\bf 109}, 174437 (2024).
\bibitem{EuPtSi_19-2} M. Kakihana, D. Aoki, A. Nakamura, F. Honda, M. Nakashima, Y. Amako, T. Takeuchi, H. Harima, M. Hedo, T. Nakama, and Y.  \={O}nuki, J. Phys. Soc. Jpn. {\bf 88}, 094705 (2019).
\bibitem{MnSc2S4_17} S. Gao, O. Zaharko, V. Tsurkan, Y. Su, J.S. White, G. S. Tucker, B. Roessli, F. Bourdarot, R. Sibille, D. Chernyshov, T. Fennell, A. Loidl, and C. R\"{u}egg, Nat. Phys. {\bf 13}, 157 (2017).
\bibitem{MnSc2S4_24} H. Takeda, M. Kawano, K. Tamura, M. Akazawa, J. Yan, T. Waki, H. Nakamura, K. Sato, Y. Narumi, M. Hagiwara, M. Yamashita, and C. Hotta, Nat. Comm. {\bf 15}, 566 (2024).
\bibitem{EuTi2Al20_Neu_21} T. Kurumaji, Y. Tokunaga, T. Arima, H. Saito, T. Nakajima, Activity Report on Neutron Scattering Research: Experimental Reports {\bf 27} (2021).
\bibitem{Kawamata_25} M. Kawamata, R. Higashinaka, T. Matsumura, M. Avdeev, K. Iwasa, H. Nakao, K. Hattori, and T. D. Matsuda, J. Phys. Soc. Jpn. {\bf 95}, 024701 (2026).
\end{thebibliography}
\end{document}